# Probing the effect of different cross sections in asymmetric collisions

Deepinder Kaur, Varinderjit Kaur and Suneel Kumar[*]
School of Physics and Materials Science, Thapar University, Patiala – 147004 (Punjab) INDIA.
*email:Suneel.kumar@thapar.edu

We present a complete systematic theoretical study of multifragmentation for asymmetric colliding nuclei for heavy-ion reactions in the energy range between 50 MeV/nucleon and 1000 MeV/nucleon by using isospin dependent quantum molecular dynamics (IQMD) model. We have observed an interesting outcome for asymmetric colliding nuclei. The comparison between the symmetric and asymmetric colliding nuclei for the isospin independent cross section and isospin dependent cross section has been studied. We have found the pronounced effect of different cross section and mass asymmetry on the nuclear reaction dynamics.

## 1. Introduction:

Heavy ion collisions offer the possibility to probe nuclear matter under different conditions of densities and temperature. At high excitation energies and temperature, the colliding nuclei may break up into many fragments known as multifragmentation[1]. In the recent times a lot of research is going on for the study of collision of mass asymmetric nuclei at intermediate energy. Multifragmentation is by essence associated to the emission of several fragments. Any study of the phenomenon requires a coincident and efficient detection of these fragments and of the associated particles ($Z \leq 2$).

With the availability of radioactive ion beam (RIB) facilities the GSI Facility for Antiproton and Ion beam Research (FAIR) [2], GANIL in France [3], RIB facility at Rikagaku Kenyusho (RIKEN) in Japan [4], at the Cooler Storage Ring (CSR) (China) [5], and the upcoming facility for RIB at Michigan State University [6], one has a opportunity to study the properties of nuclear matter under the extreme conditions.

The growing study of radioactive beam facilities at many laboratories over the world and the use of radioactive beam with large neutron or proton excess have offered an excellent tool to investigate the isospin dependence of heavy ion collision. This is helping the scientific community to obtain information about the equation of state for asymmetric nuclear matter and also the information on isospin dependence of in medium nucleon nucleon cross section, which are important not only to study nuclear reaction dynamics but also to explore the explosion mechanism of supernova and the colliding rate of neutron stars. Study of heavy-ion collisions at intermediate energies has now become important tool to investigate reaction mechanism behind various phenomenona. The study of heavy-ion collisions at intermediate energies (50 ≤E ≤1000 MeV/nucleon) provides a rich source of information for many rare phenomena such as multifragmentation, collective flow as well as particle production

The term isospin refers to the pair of similar particles, e.g. protons and neutrons, which are almost identical in the nuclear matter when electric charge difference is ignored. In many transport simulations, the nuclear interactions difference between protons and neutrons are simply ignored. In other words, these simulations explore the reactions in symmetric nuclear matter limit only [7]. The isospin effects come into the results in terms of symmetry energy and cross section. Both symmetry energy and cross section affect the multifragmentation, collective flow and related phenomena to a great extent.

The isospin effects of the in-medium NN cross section on the physical quantities arise from the difference between isospin dependent in-medium NN cross section denoted by $\sigma_{iso}$ in which $\sigma_{np} > \sigma_{nn} = \sigma_{pp}$ and isospin independent NN cross section denoted by $\sigma_{noiso}$ in which $\sigma_{np} = \sigma_{nn} = \sigma_{pp}$. Here $\sigma_{np}$, $\sigma_{nn}$ and $\sigma_{pp}$ *are* the neutron–proton, neutron–neutron and proton–proton cross sections, respectively.

For symmetric and asymmetric reactions various experimental studies offer a unique opportunity to explore the mechanism for the breaking of nuclei into pieces. At the same time, heavy ion reactions can also be used to extract the information about the nature of matter.

Jian-Ye Liu et. al. studied the isospin effects of one-body dissipation and two-body collision on the number of protons (neutrons) emitted during the nuclear reaction. Their studies show strongly that the isospin-dependent in-medium NN cross section has a much stronger influence on NP(NN) ( the number of proton(neutron) emissions)[8].

Multifragmentation has been observed both experimentally and theoretically. We will see this effect on mass asymmetric systems. The mass asymmetry of a reaction can be defined by the asymmetry parameter $\eta = | (A_T - A_P) / (A_T + A_P) |$; [9] where $A_T$ and $A_P$ are the masses of the target and projectile, respectively. The $\eta = 0$ corresponds to the symmetric reactions, whereas non-zero values of $\eta$ define different asymmetries of a reaction.

As noted by FOPI group, the reaction dynamics in a symmetric reaction ($\eta = 0$) can be quite different compared to an asymmetric reaction ($\eta \neq 0$). This is valid both at low and intermediate energies. This difference emerges due to the different deposition of the excitation energy ('in form of compressional and thermal energies) in symmetric and asymmetric reactions. The symmetric reactions lead to higher compression and asymmetric reactions lacks the compressional energy and in asymmetric reactions a large part of excitation energy is in the form of thermal energy.

As the little information is known about the in-medium NN cross section and its isospin dependence on mass asymmetry up to now, it is thus desirable to do theoretical study to gain knowledge about isospin dependence on mass asymmetry.

On the basis of theoretical scenario, one has the dynamical model where the reaction dynamics starts simulation from well defined nuclei to the end of the reaction where it is practically cold and scattered nuclear matter in the form of nucleons, light or heavy mass fragments. As a result, no dynamical model simulates the fragments; rather one has the phase space of nucleons and constructs the fragments at the end of simulations. Therefore, we look for secondary models of clusterization algorithms e.g. minimum spanning tree (MST)[18].

## 2. The model:

The dynamical model used for the present study is isospin dependent quantum molecular dynamics (IQMD) [10] model. The Isospin-dependent Quantum Molecular Dynamic model is the refinement of QMD[11] model based on event by event method. The reaction dynamics are governed by mean field, two-body collision and Pauli blocking. The IQMD [10,12] model treats different charge states of nucleons, deltas, and pions explicitly [13], as inherited from the BUU model [14]. The IQMD model has been used successfully for the analysis of a large number of observables from low to relativistic energies [10, 15]. The isospin degree of freedom enters into the calculations via the symmetry potential, cross sections, and Coulomb interactions. The details about the elastic and inelastic cross-sections for proton-proton and neutron-neutron collisions can be found in Ref. [10,16,17].

The baryons are represented by Gaussian-shaped density distributions

$$f_i(\vec{r},\vec{p},t) = \frac{1}{\pi^2\hbar^2} e^{-[\vec{r}-\vec{r}_i(t)]^2\frac{1}{2L}} e^{-[\vec{p}-\vec{p}_i(t)]^2\frac{2L}{\hbar^2}} \qquad (1)$$

Nucleons are initialized in a sphere with radius R = 1.12A$^{1/3}$ fm, in accordance with the liquid drop model. Mass dependent Gaussian width is applied. Each nucleon occupies a volume of $\hbar^3$ so that phase space is uniformly filled. The initial momenta are randomly chosen between 0 and Fermi momentum P$_F$. The nucleons of the target and projectile interact via two and three-body Skyrme forces and Yukawa potential. The isospin degrees of freedom are treated explicitly by employing a symmetry potential and explicit Coulomb forces between protons of the colliding target and projectile. This helps in achieving the correct distribution of protons and neutrons within the nucleus.

The successfully initialized nuclei are then boosted towards each other using Hamilton equations of motion

$$\frac{dr_i}{dt} = \frac{d\langle H\rangle}{d\,p_i}\;;\;\frac{dp_i}{dt} = -\frac{d\langle H\rangle}{d\,r_i} \quad (2)$$

The total Hamiltonian function with a kinetic energy T and a potential energy V is given by

$$\langle H\rangle = \langle T\rangle + \langle V\rangle$$

$$=\sum_i \frac{p_i^2}{2m_i} + \sum_i \sum_{j>i} \int f_i(\vec{r},\vec{p},t) V^{ij}(\vec{r'},\vec{r})$$

$$\times f_j(\vec{r'},\vec{p'},t)d\vec{r}d\vec{r'}d\vec{p}d\vec{p'} \quad (3)$$

The total potential is the sum of the following specific elementary potentials.

$$V = V_{Sky} + V_{Yuk} + V_{Coul} + V_{mdi} + V_{loc} \quad (4)$$

During the propagation, two nucleons are supposed to suffer a binary collision if the distance between their centroid is

$$|r_i - r_j| \leq \sqrt{\frac{\sigma_{tot}}{\pi}} \quad (5)$$

Where $\sigma_{tot} = \sigma(\sqrt{s}, type)$

The collision is blocked with a possibility

$$P_{block} = 1-(1-P_i)(1-P_j)$$

Where $P_i$ and $P_j$ are the already occupied phase space fractions by other nucleons.

The clusterization method used here is minimum spanning tree (MST)[18]. The normal MST method depends on the spatial distance and hence the fragments, thus created can have nucleons with very large relative momemta( with no momemtum cut ).

## 3. Results and discussion :

The stimulations have been carried out for three systems $^{208}Pb_{82} + ^{208}Pb_{82}$ having $\eta = 0$, $^{40}Ca_{20} + ^{208}Pb_{82}$ having $\eta = 0.6$ and $^{12}C_6 + ^{197}Au_{79}$ having $\eta = 0.8$ using IQMD model for central and semi central impact parameters at energy ranging from 50 MeV/nucleon to 1000MeV/nucleon. The soft equation of state is being used for the whole study. The phase space obtained is analyzed using minimum spanning tree (MST)[18].

Figure 1, shows the multiplicity of free nucleons and LMF's as a function of energy at scaled impact parameters. Also the effect of two cross sections one the isospin dependent cross section($\sigma_{iso}$) i.e $3\sigma_{nn} = 3\sigma_{pp} = \sigma_{np}$ and the other the isospin independent cross section($\sigma_{noiso}$) i.e $\sigma_{nn} = \sigma_{pp} = \sigma_{np}$ have been seen. One can see from the figure the difference coming out for asymmetric reactions with cross section. As the free nucleons are produced from interaction zone, there is little effect of isospin independent cross section at low energy. As the energy increases the multiplicity of free nucleons increases for isospin independent cross section.

Even for the mass asymmetric cases the trend is same for free nucleons as the trend is for symmetric cases. As it is clear from the figure that the number of free nucleons is increasing with the increase in energy. This is due to the reason that for central geometory all the nucleons are taking part in the collision. The collisions becomes more violent as the energy increases. The maximum number of free nucleons will be produced at high energy due to more compression zone produced. Also with increase in energy Pauli blocking effect decreases. The correlations among the nucleons are destroyed at high energies and hence more number of free nucleons are produced. With the increase in the value of scaled impact parameter the multiplicity of free nucleons decreases as compared to central collision. Also it would be interesting to see the effect of isospin independent cross section ($\sigma_{noiso}$). $\sigma_{noiso}$ lead to enhanced production of free nucleons at all the energies and for all asymmetries. For symmetric reactions the difference in the production of free nucleons is more for the energy range 400 MeV/nucleon to 1000 MeV/nucleon. Even for the mass asymmetric cases the trend is same for free nucleons as the trend is for symmetric cases But for asymmetric reactions, the influence is more from 200MeV/nucleon to 600MeV/nucleon

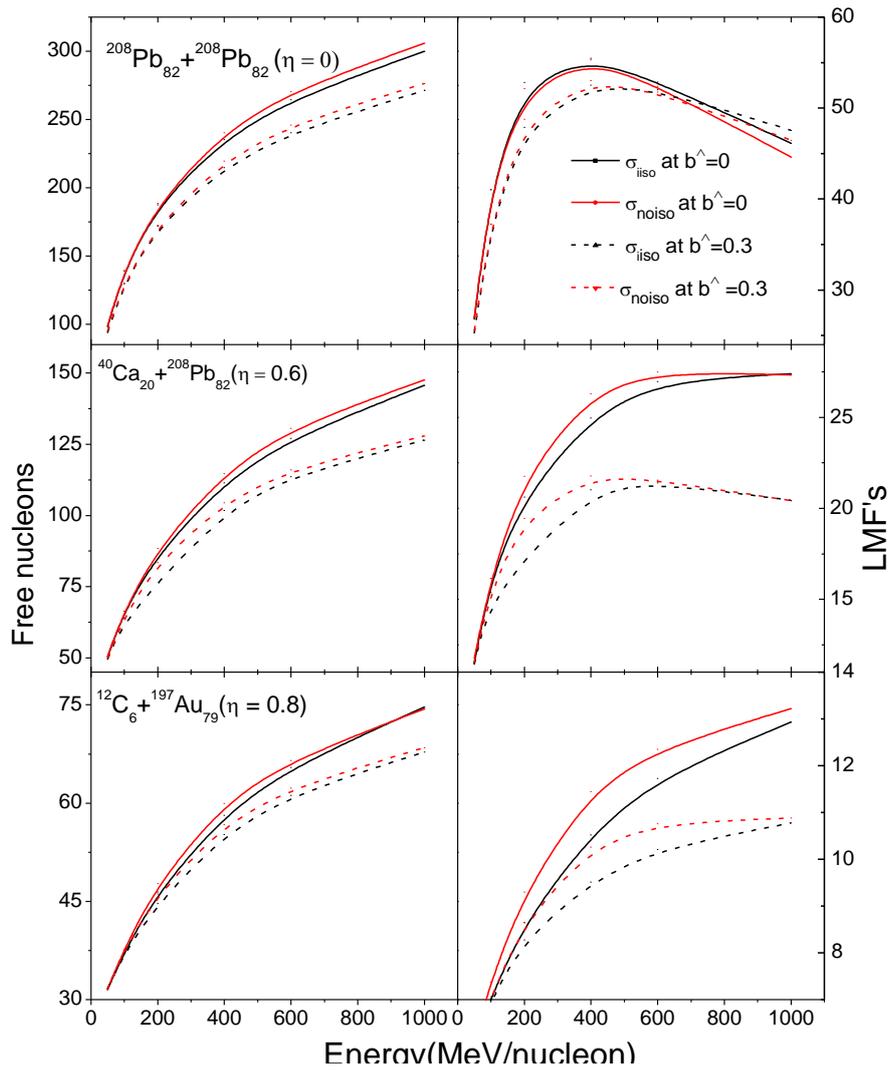

Fig.1. Multiplicity of free nucleons and LMF's as a function of energy.

The light mass fragments (LMF's) are produced from the participant zone. From the figure, it is clear that the number of LMF's first increases as energy increases, reaches a peak value at 200MeV/nucleon and then decreases as the energy increases. In mass asymmetric systems the number of LMF's increases with the increase in energy. The opposite effect that we have observed in case of free nucleons. For asymmetric systems the difference is more because mass asymmetry play significant role on reaction dynamics as studied in the ref [9].

It has been observed that there is considerable effect of cross section on the mass asymmetric systems than on symmetric systems. It can be seen from the fig.1 that the number of light mass fragments formed without isospin dependent nucleon nucleon cross section is more as compared to symmetric systems. In case of symmetric systems the number of LMF's is decreasing with increasing energy for independent cross section.

Figure 2, shows the multiplicity of intermediate mass fragments as a function of energy for symmetric and asymmetric reactions. As the asymmetry of the reaction increases, the trend of rise and fall in the multiplicity of IMF's is not followed.

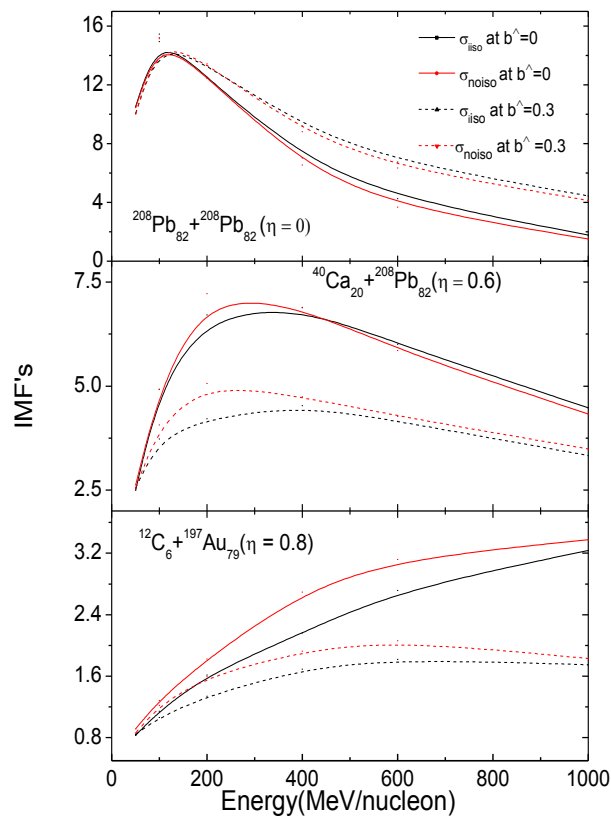

Fig. 2. Multiplicity of IMF's as a function of energy.

The effect of cross section is more for free nucleons and LMF's as compared to IMF's because IMF's are produced from spectator zone. It is seen that multiplicity of intermediate and light mass fragments decrease with the increase in energy but in contrary it goes on increasing with energy for free nucleons. The emission of free nucleons will show disassembly (vaporization) of the matter [19,20]. The production of LMF's is highest at low energy which decreases with the increase in energy. Due to very small overlap at large impact parameter, the system does not receive enough energy and hence cools down after emitting few nucleons/LMF. The production of IMF's is maximum at low energy and largest impact parameter.

In figure 3, the multiplicity of various fragments is displayed as a function of total mass of the system at time 200fm/c at energy ranging from 200 MeV/nucleon to 1000 MeV/nucleon. The total mass of these reactions $^{208}Pb_{82}$ + $^{208}Pb_{82}$ (416) having η = 0. $^{40}Ca_{20}$ + $^{208}Pb_{82}$ (248) having η = 0.6 and $^{12}C_6$ + $^{197}Au_{79}$ (209) having η = 0.8 have been displayed. The universal behavior of increase in multiplicity of fragments with the size of the system is observed in the presence of mass asymmetry as well as with isospin independent cross section. One can see from the figure that the trend for the isospin dependent cross section and for the isospin independent cross section is same.

Both free nucleons and LMF's show increasing trends. With the increase in the size of system, number of the participant nucleons increases. This will lead to more thermalization of the system. Due to this reason, increase in multiplicity of fragments will always be observed, which will originate from the participant zone.

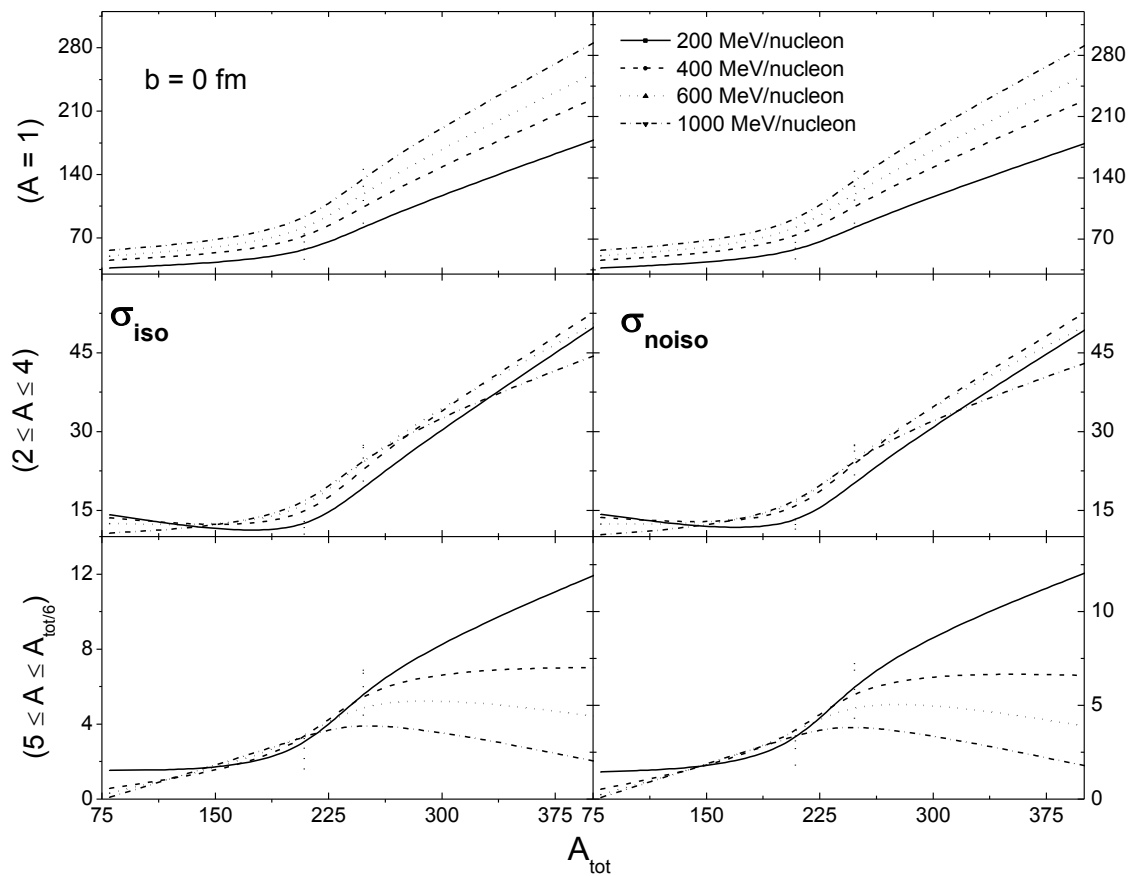

Fig.3. Multiplicity of various fragments vs total mass of the system($A_{tot}$) at different energies

In the figure 4, the multiplicity of intermediate mass fragments as a function of impact parameter (b) has been displayed at 600MeV/nucleon for C + Au system. The system is having asymmetry equal to 0.8. From the figure it can be seen that the maximum value of IMF is obtained for lower values of b. As the value of impact parameter increases, the mean value of IMF multiplicity is decreases. Calculations has been compared with experimental data of Aladin group [21]. The open circles in the figure shows the experimental data and the solid squares show the calculations using IQMD model. It has been seen that the calculations are corresponding to the experimental data available. The trend of theoretical result follow the experimental data.

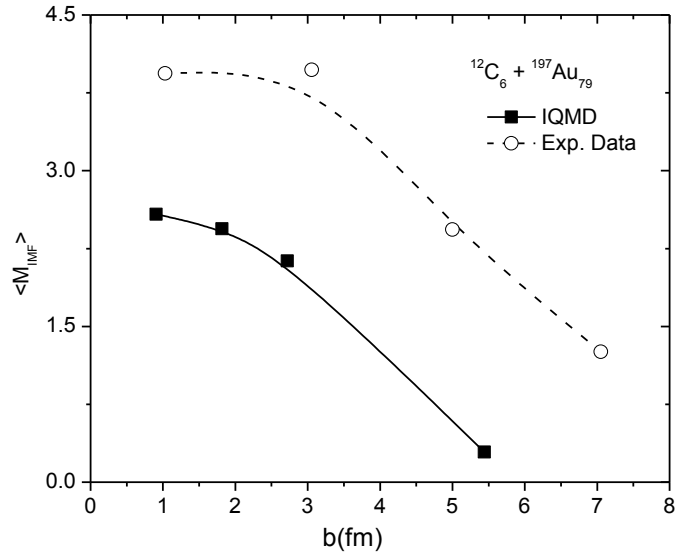

Fig. 4. Mean IMF multiplicity vs impact parameter b, for the experimental data and the IQMD calculations for $^{12}C_6 + ^{197}Au_{79}$ at 600MeV/nucleon.

## 4. Summary:

We present a complete systematic theoretical study of multifragmentation for asymmetric colliding nuclei for heavy-ion reactions in the energy range between 50 MeV/nucleon and 1000 MeV/nucleon using IQMD model. We envision an interesting outcome for asymmetric colliding nuclei. The effect of isospin independent cross section and isospin dependent cross section has been studied for different mass asymmetries. We have found the pronounced effect of different cross section and mass asymmetry on the nuclear reaction dynamics. A similar trend is observed between the theoretical calculations and the experimental data of Aladin.

## Acknowledgements:

This work has been supported by the grant from Department of Science and Technology (DST), Government of India, vide Grant No.SR/WOS-A/PS-10/2008.